\documentclass[twocolumn,showpacs,preprintnumbers,amsmath,amssymb,eps]{revtex4}
\usepackage{graphicx}% Include figure files
\usepackage{dcolumn}% Align table columns on decimal point
\usepackage{bm}% bold math

\begin{document}

\title{Phase Separation and the Phase Diagram in Cuprates Superconductors}
\author{E. V. L. de Mello}
\altaffiliation[]{evandro@if.uff.br}
\author{D. H. N. Dias }
\affiliation{
Instituto de F\'{\i}sica, Universidade Federal Fluminense, Niter\'oi, RJ 24210-340, Brazil\\}
\author{Otton Teixeira da Silveira Filho }
\affiliation{
Instituto de Computa\c{c}\~ado, Universidade Federal Fluminense, Niter\'oi, RJ 24210-340, Brazil\\}%
\date{\today}
\begin{abstract}

We show that the main features of the cuprates superconductors phase
diagram can be derived considering the disorder as a key property of
these materials. Our basic point is that the high pseudogap line is
an onset of phase separation which generates compounds made up of
regions with distinct doping levels. We calculate how this
continuous temperature dependent phase separation process occurs in
high critical temperature superconductors (HTSC) using the
Cahn-Hilliard approach, originally applied to study alloys. Since
the level of phase separation varies for different cuprates, it is
possible that different systems with average doping level $p_m$
exhibit different degrees of charge and spin segregation.
Calculations on inhomogeneous charge distributions in form of
stripes in finite clusters performed by the Bogoliubov-deGennes
superconducting approach yield good agreement to the pseudogap
temperature $T^*(p_m)$, the onset of local pairing amplitudes with
phase locked and concomitantly, how they develop at low temperatures
into the superconducting phase at $T_c(p_m)$ by percolation.
\end{abstract}


\maketitle

\section{Introduction}

After almost twenty years of research, the high critical temperature
superconductors (HTSC) still remains as an unsolved
problem\cite{Nature,TS,Tallon}. All HTSC  have a similar universal
complex electronic phase diagram: the parent (undoped) compound is
an antiferromagnetic Mott insulator, the superconducting phase has a
dome shape at low doping and low temperature. The normal phase has a
pseudogap in the underdoped region at low temperatures and a
metallic phase at high temperatures  in the overdoped region.
Understanding the complexity of such normal phase is believed to be
essential for solving the mechanism of HTSC\cite{Nature}.

Another intriguing fact
is the question of the inhomogeneities in these materials;
some families of compounds have
a high inhomogeneous electronic structure which display
either stripe\cite{Tranquada}, patchwork\cite{Pan}, checkerboard\cite{Hanaguri},
or other forms\cite{Lee}. On the other hand, depending
on the type of experiment, there are some HTSC materials that appear to
be more homogeneous or, at least do not display any  gross
inhomogeneity\cite{Bobroff,Loram}. It is possible that these
distinct features are due to a phase separation
transition that produces different degrees of local hole doping
densities and consequently, different properties.
We have already discussed this possibility in a previous
paper\cite{Mello04} and we now perform more detailed calculations
in connections with several new experimental data.

Recent angle resolved photoemission (ARPES) experiments with
improved energy and momentum  resolution\cite{DHS,Zhou,Zhou04,Ino}
have distinguished two  electronic components in $\vec k$-space
associated with the $La_{2-x}Sr_xCuO_4$ (LSCO) system: a metallic
quasi particle spectral weight at the ($\pi/2,\pi/2$) nodal
direction which increases with hole doping and an insulator like
spectral weight at the  end of the Brillouin zone straight segments
in the ($\pi,0$) and (0,$\pi$) antinodal regions which are almost
incentive to the doping level. Comparison with the non
superconducting $La_{2-x-y}Nd_ySr_xCuO_4$ system\cite{Zhou}, in
which static stripes were first observed\cite{Tranquada},
demonstrated that the antinodal spectral weight behavior is
compatible with a quasi one dimensional electronic structure where
the hole rich stripes behaves as one dimensional metals and the hole
poor stripes as insulators. These features demonstrated that in
these compounds there are two aspects of the electronic
structure\cite{DHS,Zhou,Zhou04,Ino}. Moreover, the large shift of
the ARPES spectra\cite{DHS} at the Fermi energy, what is called the
{\it leading edge shift} and it is interpreted as the
superconducting gap is maximum at the antinodal region. This is an
indication of the d-wave symmetry of the superconducting order
parameter, the zero temperature superconducting gap $\Delta_0$
vanishes in the nodal but it is maximum in the antinodal directions.

Another technique which has been refined and revealed new aspects of HTSC is
scanning tunneling microscopy (STM). It is complementary to ARPES
because it probes the differential
conductance or the pairing amplitude $\Delta$ directly on the surface of the compound.
New STM data with great resolution have also revealed strong inhomogeneities in
the form of a patchwork of
(nanoscale) local spatial variations in the density of states which is
related to the local superconducting gap\cite{Fournier,Pan,Davis}.
More recently it was possible to distinguish two distinct
behavior: well defined coherent
and ill-defined incoherent peaks depending on the exactly spectra location
at a $Bi_2Sr2CaCu_2O_{8+\delta}$ (Bi2212) surface\cite{Hoffman,McElroy1,Fang}.
STM experiments have also detected a regular
low energy checkerboard order in the electronic structure of
the Bi2212 family above the superconducting
critical temperature ($T_c$)\cite{Vershinin} and at low
temperature\cite{McElroy2} and in the $Na_xCa_{2-x}CuO_2Cl_2$\cite{Hanaguri}.

A third important set of experiments to describe the HTSC phase
diagram is the tunneling current\cite{Renner,Miyakawa,Suzuki}. New
techniques have recently shown the existence of two energy gaps
which behaved differently under an applied magnetic
field\cite{Deutscher,Krasnov,Yurgens}. Tunneling experiments using
superconductor insulator superconductor (SIS) with insulator layers
with various size and resistivity have also shown distinct sets of
energy scales and have also led to the idea that the richness of the
phase diagram as function of doping may be due to charge
inhomogeneities and charge cluster of different size in the Cu-O
planes\cite{Moura,Mourachkine}.

The ARPES and STM experiments, are surface probes
what may suggest that the inhomogeneities
may be a surface effect, but charge disorders were also detected
by bulk experiments, like the stripe phase in materials similar
to LSCO\cite{Tranquada,Bianconi} and also local variations in the charge
\cite{Bozin}. Another bulk sensitive experiment is
nuclear magnetic and quadrupole resonance (NMR and NQR) which have
provided ample evidence for spatial charge inhomogeneity in the $CuO_2$
planes\cite{Curro,Singer,Haase}. Similarly, Singer at al\cite{Singer}
measured a distribution of $T_1$ over the Cu NQR spectrum in bulk
LSCO which can be attributed to a  distribution of holes $p$ with
a half width of $\Delta p/p\approx 0.5$. More recently, NMR results
on $La_{1.8-x}Eu_{0.2}SrCuO_{4}$ were interpreted as evidence for
a spatially inhomogeneous charge distribution in a system where the
spin fluctuations are suppressed\cite{NCurro}. This new result is
a strong indication
that the charge disorder may be due to a phase separation transition.
%It is likely that the origin
%of the charge stripes may be due to an anisotropic mesoscopic
%Jahn-Teller interaction
%between electrons and optical phonons\cite{Muller,Kabanov}.

These unusual features of cuprates led to theoretical proposals that
phase separation is essential to understand their physics\cite{Zaanen,Machida}.
In fact, phase segregation has been observed on the $La_2CuO_{4+\delta}$ by
x-ray and transport measurements\cite{Grenier,Jorg}. They have
measured a spinodal phase segregation into an oxygen-rich (or hole-rich)
metallic phase and an oxygen-poor antiferromagnetic phase above T=220K.
Below this temperature the mobility of the interstitial oxygen becomes
too low for a further segregation. $La_2CuO_{4+\delta}$  is the only
system where ion diffusion has been firmly established, although there are
evidence of ion diffusion at room temperature in micro crystals of the
Bi2212 superconductors at a very slow rate\cite{Truccato}.

On the other hand, recent NMR studies on $YBa_2Cu_3O_{6+y}$ (YBCO) have demonstrated
a complete absence of static phase separation or at least an absence
of gross inhomogeneities\cite{Bobroff,Loram}. Contrary to what was measured
in Bi2212 and in LSCO, the maximum hole doping variation $\Delta p$
found in YBCO is very small\cite{Bobroff}. Loram et al\cite{Loram} analyzed
also the specific heat of YBCO and Bi2212 and concluded that there are evidences
for an uniform doping density in these materials. Surface measurements are
difficult to be performed in YBCO and therefore they are not as conclusive
as in Bi2212 or LSCO\cite{DHS,Bobroff}.

In this paper we develop the idea of two distinct\cite{Kris} pseudogaps in
which the lower one is associated with the onset of
superconductivity\cite{Markiewicz}. We then take the upper pseudogap
line as a line of phase separation transition, and introduce a model
to make quantitative predictions through the Cahn-Hilliard (CH)
approach\cite{Mello04,CH,Otton} in section II. In this way, gross
and weak disordered systems differ only by the degree of mobility or
diffusion of the particles and  different systems can belong to the
same universality class. In section III we use the
Bogoliubov-deGennes (BdG) local method to the superconducting
problem to calculate the local pairing amplitude in mesoscopic
clusters with random, stripes, checkerboard, Gaussian and other
forms of inhomogeneities in the charge density. The results on
stripe-like formations applied to the LSCO system reveal a pseudogap
phase characterized by the building up of superconducting islands or
puddles with, as in BCS theory, the phase locked. Consequently the
superconducting phase is reached at low temperatures by the
percolation of these islands\cite{OWK,Mello03,JL}. The details of these
calculations are discussed in section  II and III with their
consequences to HTSC in section IV.

\section{The CH Approach }

 Our main assumption is that the high pseudogap line, which we call $T_{ps}(p_m)$,
falls to zero near $p\approx 0.20$, and is independent of the
superconducting phase\cite{Tallon,Naqib}, is the onset of the phase
separation. Timusk and Sttat called this line a crossover
boundary\cite{TS} and it is distinct of the lower
pseudogap\cite{Markiewicz}. Thus a given system start to phase
separate at $T_{ps}(p_m)$ and this process increases continuously as
the temperature goes down.

Thus the phase separation problem in HTSC
is a dynamical and depends strongly on the initial conditions, on
the temperature and how the system is quenched below the phase separation
line, the mobility of hole and ions, and so on. However, the
information on most of these procedures are not available and on has to work out
the phase separation process backwards, to wit, use parameters
which yield the final configurations of
stripe, checkerboard, or other  patterns.
An appropriate framework to
study such process mathematically is by the CH theory\cite{CH},
which we have already applied to the cuprates\cite{Mello04}.

The CH approach to phase separation was conceived to describe the
continuous transition of binary alloys, but, in principle, can be
applied to any system that undergoes this type of continuous
transition\cite{CH}. As we can infer from the stripe phases, in a
compound with an average of $p_m=1/8$ hole per copper atom, the
antiferromagnetic insulating phase has stripes of nearly zero holes
per copper atom and metallic ones with larger values of local charge
density. As we discuss below, this behavior can be well described by
the CH theory.

Starting with small fluctuations of the local charge density,
the CH non-linear differential equation which describes the phase
separation at a temperature $T$ below
the phase separation transition at $T_{ps}$ can be written as\cite{CH,Mello04}:

\begin{eqnarray}
\frac{\partial u}{\partial t} = -M\nabla^2(\varepsilon^2\nabla^2u
+ A^2(T)u-B^2u^3),
\label{EqCH}
\end{eqnarray}
where $u$ is the order parameter associated with the local variation $p(\vec x)$
in the  average number of holes per copper atom $p_m$ at a given point $\vec x$,
defined as $u(\vec x)\equiv p(\vec x)-p_m$ and we expected
$u(\vec x) \approx 0$ above and near the $T_{ps}$. $\varepsilon$,
and $B$ are fixed parameter, the parameter $A(T)$
depends on the temperature $T$ and $B$ and
the ratio $\pm A(T)/B$ yields the two equilibrium densities. $M$ is the mobility
of the particles and dictates the time scale of the phase separation process. Compounds
with larger values of $M$ phase separate easier than those with smaller values,
and, therefore, it will differentiate among the different HTSC families.
As the temperature
goes down below $T_{ps}(p_m)$, the two equilibrium order parameter
(or densities) spread
apart from one another and the energy barrier
between the two equilibrium phases $E_g(T,p_m)$ also increases. $E_g=A^4(T)/B$
which is proportional to $(T_{ps}-T)^2$. $E_g$ can be identified with the
upper pseudogap signal\cite{Mello04} which is detected by several
different experiments\cite{Tallon,Markiewicz}.
A discussion on the details of the
convergence criteria, boundary conditions and dependence
on the initial parameters in one, two or three dimensions,
can be found in our previous work\cite{Otton}.

\begin{figure}[!ht]
%\begin{center}
\includegraphics[height=7cm]{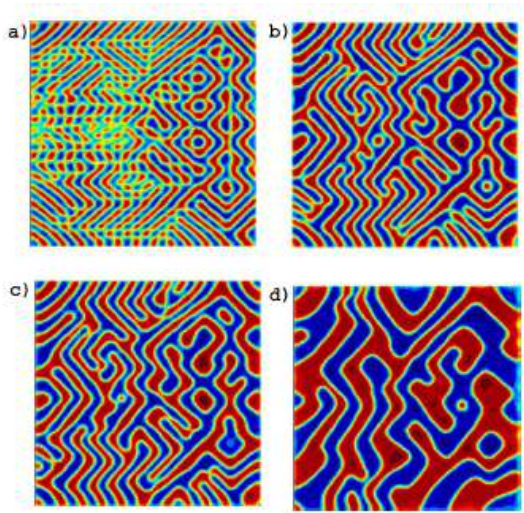}
%\end{center}
\caption{ (color online) The mapping of the density order parameter during  the  process of
phase separation. The start order parameter is small random variations
around $u=0$. The order of the figures is a) t=400 time steps, b) t=800,
c) t=1000 and d) t=4000.}
\label{1801ac}
\end{figure}

\begin{figure}[!ht]
%\begin{center}
\includegraphics[height=7cm]{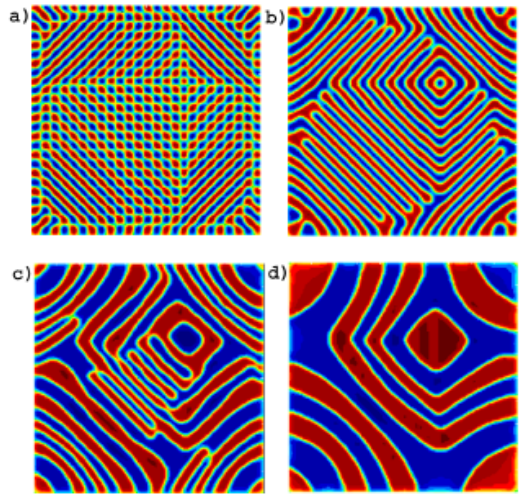}
%\end{center}
\caption{ (color online) The same time evolution of Fig.1 but with a
more symmetric
initial condition around $u=0$. The order of the
figures is a) t=400 time steps, b) t=1000,
c) t=4000 and d) t=20000. For shorter times, the phase
separation process develops  symmetry patterns which are lost
for larger times.}
\label{1801s}
\end{figure}

In Figs.(\ref{1801ac}) and (\ref{1801s}) we display the mapping of
the order parameter for a system with $p_m=1/8$ and with a maximum
$\Delta p=1/8$. Both figures have a small initial density
fluctuation of $\Delta p=0.02$ but Fig.(\ref{1801ac}) has more
random initial conditions than the more symmetric case of
Fig.(\ref{1801s}). We notice that, for larger times they are almost
indistinguishable but for short times they are very different. It is
interesting that, despite the large time evolution, the system keeps
very symmetric patterns. In Fig.(\ref{1801s}a) we can see a mixture
of checkerboard and stripe formations  which tends to evolve into a
pure stripe phase (Fig.(2b)). As the phase separation process
continues, the systems tends to a complete phase separation and the
early symmetric patterns are lost. This degree of charge phase
separation can be reached very fast if the holes have a large
mobility $M$. This finding enable us to speculate that the
differences measured among the HTSC families are due to  different
values of $M$. By the same token, the differences found in compounds
of a given family can be attributed to differences in the quenched
process, that is, the rate of cooling below $T_{ps}(p)$.

\begin{figure}[!ht]
%\begin{center}
\includegraphics[height=7cm]{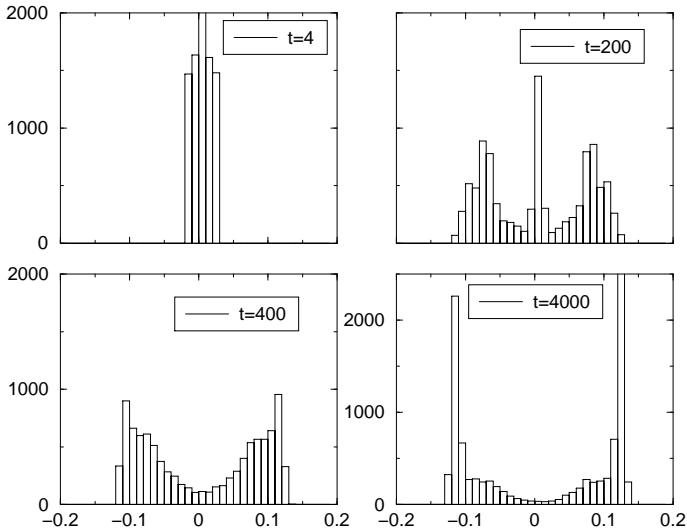}
%\end{center}
\caption{ The histogram time evolution of the density order
parameter of Fig.1  with a symmetric
initial condition around $u=0$.}
\label{hevol1801s}
\end{figure}

Another possible way to follow the whole phase separation process is
to analyze the charge histogram evolution with time. Thus, in
Fig.(\ref{hevol1801s}) we show the phase separation progress in
terms of the histograms of the order parameter. One can see the
tendency, as the time flows, to form a density pattern of a bimodal
distribution around the two equilibria conditions $p(i)_{\pm}=\pm
A/B$. Thus, systems with a high mobility will probably reach a state
where there are two type of regions with high and low densities.
This is similar to the stripe phase in the LSCO system.

Therefore, taking the large pseudogap\cite{Tallon,Mello04,Naqib}
as the phase separation temperature $T_{ps}$, it is possible to
infer that the different HTSC unfold in different patterns as
described by Figs. 1 and 2. As a consequence, since underdoped compounds have
a very high $T_{ps}$, they phase separate into a bimodal distribution
faster than the optimally and overdoped compounds. The
NQR results of Singer et al\cite{Singer} suggest a bimodal
distribution of charge in the LSCO system.

Thus, in order to calculate the local critical temperature for such
an inhomogeneous system, we need to use the Bogoliubov-deGennes
theory. This will be outlined in the next section.

\section{The Local Superconducting Calculations}

Here we discuss the main points of a local
superconducting calculation  to deal with the effect of the
charge disorder which follows directly from the CH patterns described above.
The general way to perform this, in
a system without spatial invariance, is through the
BdG mean-field theory which has been largely used in the
HTSC problem\cite{Franz,Ghosal,Ghosal2,Nunner}.
The important and novel point introduced  here is that we take the
initial charge distribution derived from the CH results.
The procedure starts with
the extend Hubbard Hamiltonian

\begin{eqnarray}
&&H=-\sum_{\ll ij\gg \sigma}t_{ij}c_{i\sigma}^\dag c_{j\sigma}
+\sum_{i\sigma}(\mu_i)n_{i\sigma}  \nonumber \\
&&+U\sum_{i}n_{i\uparrow}n_{i\downarrow}+{V\over 2}\sum_{\langle ij\rangle \sigma
\sigma^{\prime}}n_{i\sigma}n_{j\sigma^{\prime}}
\label{Hext}
\end{eqnarray}
where $c_{i\sigma}^\dag (c_{i\sigma})$ is the usual fermionic creation (annihilation)
operators at site ${\bf x}_i$,
spin $\sigma \lbrace\uparrow\downarrow\rbrace$, and
$n_{i\sigma} =  c_{i\sigma}^\dag c_{i\sigma}$.
$t_{ij}$ is the  hopping between site $i$ and $j$. Here we have
implemented in our calculations
hopping values up to $5^{th}$ neighbors derived from the ARPES data
for YBCO\cite{Schabel}.
In their notation,
the hopping parameters are:
$t\equiv t_1$=0.225eV, $t_2$/$t_1$=-0.70,
$t_3$/$t_1$=0.25, $t_4$/$t_1$=0.08, $t_5$/$t_1$=-0.08.
$U=1.1t$ is the on-site and  $V=-0.6t$ is the nearest neighbor
phenomenological interactions.
$\mu_i$ is the local chemical potential. All the calculations
presented here use  the same set of parameters, and
clusters with periodic boundary conditions.

\begin{equation}
\begin{pmatrix} K         &      \Delta  \cr\cr
           \Delta^*    &       -K^*
\end{pmatrix}
\begin{pmatrix} u_n({\bf x_i})      \cr\cr
                v_n({\bf x_i})
\end{pmatrix}=E_n
\begin{pmatrix} u_n({\bf x_i})       \cr\cr .
                 v_n({\bf x_i})
\end{pmatrix}
\label{matrix}
\end{equation}

These BdG equations
are solved self-consistently together with the pairing
amplitude\cite{Franz}
\begin{eqnarray}
\Delta_U({\bf x}_i)&=&-U\sum_{n}u_n({\bf x}_i)v_n^*({\bf x}_i)\tanh{E_n\over 2k_BT} ,
\label{DeltaU}
\end{eqnarray}

\begin{eqnarray}
\Delta_{\delta}({\bf x}_i)&=&-{V\over 2}\sum_n[u_n({\bf x}_i)v_n^*({\bf x}_i+{\bf \delta})
+v_n^*({\bf x}_i)u_n({\bf x}_i  \nonumber \\
&&+{\bf \delta})]\tanh{E_n\over 2k_BT} ,
\label{DeltaV}
\end{eqnarray}
and the hole density is given by
\begin{eqnarray}
p({\bf x}_i)=1-2\sum_n[|u_n({\bf x}_i)|^2f_n+|v_n({\bf x}_i)|^2(1-f_n)],
\label{density}
\end{eqnarray}
where $f_n$ is the Fermi function.

We have performed self consistent calculations with
Eqs.(\ref{DeltaV}) and (\ref{density}) on clusters up to $24\times
24$ sites with homogeneous and inhomogeneous local doping. The major
difference from previous BdG calculations is that instead of an
impurity potential\cite{Ghosal,Ghosal2} to account for the charge
disorder, we have fixed the initial local charge densities
throughout the calculations in order to take into account the
results of the CH approach and also in agreement with the data on
stripe formation\cite{Tranquada,Bianconi}. This novel procedure is
necessary to study the formation of the superconducting regions on
the local density patterns which results from the phase separation
process as shown, for instance, in Figs.1 and 2. In the next
section, we perform the superconducting calculations with d-wave
symmetry for clusters with disordered local charge.

\section{Results}

\begin{figure}[!ht]
\begin{center}
\includegraphics[height=2cm]{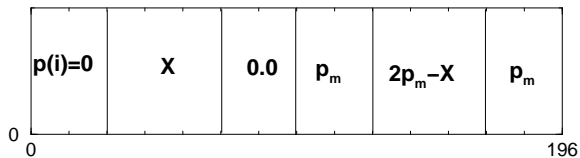}
\caption{ The low temperature stripe profile of a
$14 \times 14$ cluster  with 196 sites used to model a  compound of average
doping level of $p_m$. The BdG superconducting calculations are made
on these clusters and $0 \le X \le 0.06$
measure deviations from a bimodal charge distribution.}
\label{cluster}
\end{center}
\end{figure}

We start with BdG calculations on clusters of uniform density
ranging from zero to $p_m=0.3$ with the hopping parameters which
make the values of $T_c(p_m)$ vanish near these limits as listed
below Eq.2. Then we perform calculations with inhomogeneous clusters
and concentrate on the stripe geometry  which occur in
LSCO\cite{Tranquada}. We connect the stripe phase with the above CH
results through a scheme displayed in Fig.(\ref{cluster}) where the
value of $X$ is related with the degree of the phase separation. Due
to the large values of $T_{ps}(p_m)$ at the strong underdoped
regime, the level of phase separation is maximum at low
temperatures, and $X=0$ for doping values of $p_m \le 0.05$. For
$0.12\le p_m \le 0.19$ the value of X may build up to 0.05. Notice
that when X=0, the charge distribution is a bimodal and the system
is divided into two distinct regions; insulator and metallic. In
this case the metallic regions are in the limit of
percolation\cite{Stauffer}, but for X=0.04 or bigger, despite of the
insulator regions present, the metallic character dominates over the
entire system. This scheme is an approximate way to deal with the
phase separation which leads to the stripe charge configuration of
real systems.

\begin{figure}[!ht]
%\begin{center}
\includegraphics[height=6cm]{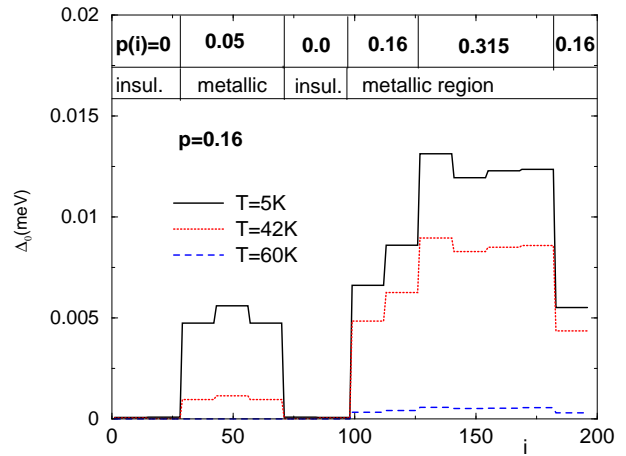}
\caption{ (color online) The temperature evolution of the local
pairing gap amplitude $\Delta (i,T)$ (in units of $eV$) for systems
with stripe disorder and average doping level of $p_m=0.16$ with
$T^*\approx65$K and $T_c\approx42$K.} \label{p016}
%\end{center}
\end{figure}

\begin{figure}[!ht]
%\begin{center}
\includegraphics[height=7cm]{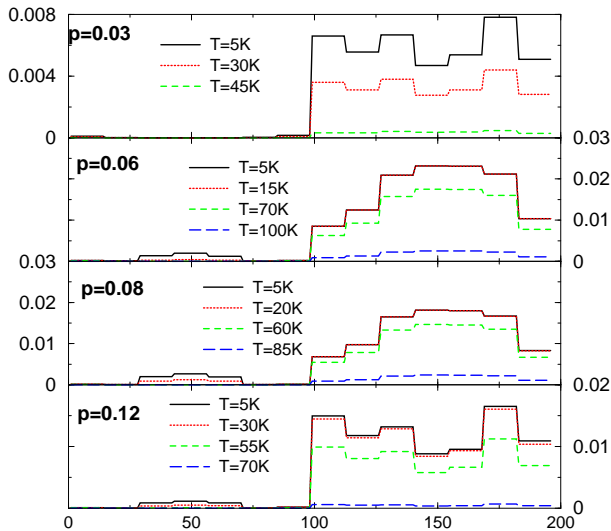}
\caption{ (color online) The temperature evolution of the local
pairing amplitude $\Delta (i,T)$ (in units of $eV$) for systems
with stripe charges with average doping level ranging from
$p_m=0.03$ to $p_m= 0.12$. The onset temperature ($T_c$) of
percolation is shown in every  panel}
\label{Deltan03.12T}
%\end{center}
\end{figure}

Thus, the goal is to study the {\it local} pairing gap at a site
${\bf x_i}$ or simply "i" as function of temperature  $\Delta (i,T)$
on clusters with charge stripes. Following the values of $\Delta
(i,T)$ it is possible to draw many  interesting consequences to the
phase diagram. To explain our approach, we will firstly analyze the
results for a cluster with $p_m=0.16$ in a $14 \times 14$ cluster as
shown in Fig.(\ref{p016}). At high temperatures but below
$T_{ps}(p_m=0.16) \approx 400$K\cite{Tallon}, the system is a
disordered metal going through a continuous phase separation as the
temperature is decreased. At low temperatures ($\le 200K$) this
compound may be composed of six stripes with local charge density
$p(i)$ given by $0.0-0.05-0.0-0.16-(0.32-0.05)-0.16$. The low
density regions are at the left and the high density are at the right of the
cluster and one follows the other by periodic boundary conditions.
These two markedly different regions were clearly detected by
several ARPES measurements\cite{DHS,Zhou,Zhou04,Ino}. At high
temperatures there is no superconducting regions in the sample, but
at $T^*=65$K, as we can see in Fig.(\ref{p016}), some local
superconducting gap arises forming  superconducting islands at the
most dense or metallic regions. Thus this temperature is the onset
of pair formation or the beginning of the pseudogap phase but these
superconducting regions are isolated from one another in the
disordered metallic matrix\cite{OWK,Mello03,JL}. Upon cooling down, the
superconducting regions become more robust and new ones are built up
at the metallic  and also where there is a hole fluctuation (at $X$
in Fig.(\ref{cluster})) at the lightly doped regions. The induced
pairing amplitude  at the insulator regions is an unexpected phenomena
and it is remarkable that it is  the origin of the
superconducting phase through the percolation of the local superconducting
order parameter at different locations
with their phase locked. As seem in Fig.(\ref{p016}), around 42K the
pairing amplitudes develop also at $X=p(i)=0.05$ but for the $p_m=0.06$ sample,
they develop even at $X=p(i)=0.001$ (see Fig.(\ref{Deltan03.12T})) and thus, the
superconducting regions cover more than 50\% of the $CuO_2$ plane.
Consequently they percolate\cite{Stauffer} and the whole system  is able
to hold a current without dissipation, that is, $T_c(p_m=0.16)=42$K
is the superconducting critical temperature for this compound.

Notice that the assumption that the pairing amplitudes have a rigid
phase, as in a BCS superconductor, has also
experimental support\cite{Bergeal}, although it is against the
phase disordered scenario which lately has gained increased theoretical
interest\cite{Lee,Emery}.

\begin{figure}[!ht]
%\begin{center}
\includegraphics[height=7cm]{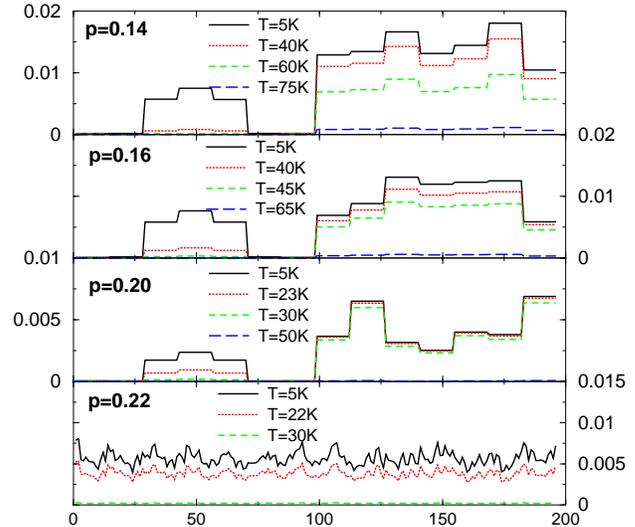}
\caption{ (color online) The temperature evolution of the local
pairing gap amplitude $\Delta (i,T)$ (in units of $eV$) for systems
with stripe disorder with average doping level ranging from
$p_m=0.14$ to $p_m= 0.22$. Notice that the $p_m= 0.22$ compound does
not undergoes a phase separation transition and its density
fluctuates around $p_m= 0.22$.} \label{Deltan14.22T}
%\end{center}
\end{figure}

We now apply this analysis to a series of compounds in order to show
how the main features of the whole LSCO phase diagram can be
derived. Our results are shown in Fig.(\ref{Deltan03.12T}) for
mostly underdoped samples and in Fig.(\ref{Deltan14.22T}) for larger
average doping values. For very lightly doped compounds like
$p_m=0.03$, due to the high values of $T_{ps}(p_m)$, there are
stripes of only $p(i)=0.0$ and $p(i)=0.06$ separated by a small
boundary with $p(i)=0.03$. We label a region as metallic if it has a
density of $p(i)\ge 0.05$, since the doped regions with $p(i)=0.06$
occupy less than half system, what is lower than the two dimensional
percolation limit of 50-60\%\cite{Stauffer}, this compound is not a
metal, although it has a metallic behavior at high
temperatures\cite{Yoshida}. Such behavior can be explained as due to
the holes which can tunnel over the dense $p(i)=0.06$ regions. This
tunneling can also be the origin of the zero temperature pseudo gap (ZTPG) detected by
STM\cite{McElroy1,Fang,McElroy2}. For these underdoped compounds, the
pairing amplitude develops strictly in these metallic or heavy
doped regions (see top panel of Fig.(\ref{Deltan14.22T})), and the
superconducting islands occupy less than $50\%$ of the available
area, what is below the percolation limit.

For a compound with $p_m=0.06$,  $T_{ps}$ is still very high, the
density profile is characterized by a very small charge fluctuations
around a bimodal distribution given by $X=0.001$. The stripe regions
have densities given by $p(i)=0.0-0.001-0.0-0.06-0.119-0.06$. This
system can be considered in the metallic limit which is $50\%$ of
the sites with $p(i)\ge0.04$. We see that the onset of
superconducting islands is at $T^*=100$K and the percolation
threshold occurs at $T_c=15$K, specially due to the already mentioned
unexpected pairing amplitudes induced at the low density sites with
$p(i)=0.001$ (see Fig.(\ref{Deltan03.12T})). Thus for temperatures
between $T=15$K and $T=100$K the system is a poor metal with
insulator, metallic and superconducting regions. The presence of
these superconducting islands in many compounds are verified by
several different experiments. Perhaps the most clear demonstration
of these static superconducting regions is through the tunneling
data\cite{Renner,Mourachkine}. More recently, measurements of
the Nernst effect\cite{Ong} demonstrated also the presence of the
local superconducting regions above $T_c(p_m)$ although it was
interpreted as due to superconducting fluctuations instead of the
static cluster considered here. For compounds with an average doping
larger than $p_m=0.06$, we notice that the onset of
superconductivity $T^*$ decreases almost steadily, while the onset
of percolation $T_c$ goes through a maximum at the optimum doping
$p_m=0.16$. The reason for this behavior in our calculation is
the deviations from the bimodal distribution given by the increase
of the phase separation parameter $X$.

In the Fig.(\ref{Deltan14.22T}) we can see that the phase separation
process and the local superconducting calculations follow a similar
pattern up to $p_m=0.20$. Following current
trends\cite{Tallon,Naqib,Markiewicz}, the phase separation ends at
$p_m\approx 0.20$ and, for heavier doped compounds, the charge
disorder is very weak like a small fluctuation around the average
value $p_m$, similar to the charge distribution shown in Fig.(3a).
This is the same kind of charge fluctuation which occurs above the
phase separation transition temperature $T_{ps}$. Consequently, the
compound with $p_m=0.22$ differs greatly from the others compounds
shown in Fig.(\ref{Deltan14.22T}), and it has only local densities
$p(i)\approx 0.22$ which is in the metallic range. The pairing gaps
$\Delta(i,T)$ are built in the whole system at $T=30$K and not in
islands or droplets. If we adhere to the assumption that the $T^*$
is the onset of superconducting correlations, we see that, for these
$p_m\ge 0.20$ overdoped samples $T^*$ merge into $T_c$. The other
important consequence is that the normal phase is much more
homogeneous, without insulating and superconducting regions, than
the $p_m \le 0.20$ compounds. This is seem experimentally through
the Fermi liquid behavior of many measurements carried in the heavy
overdoped regime\cite{TS,Lee}.

\begin{figure}[!ht]
\includegraphics[height=7cm]{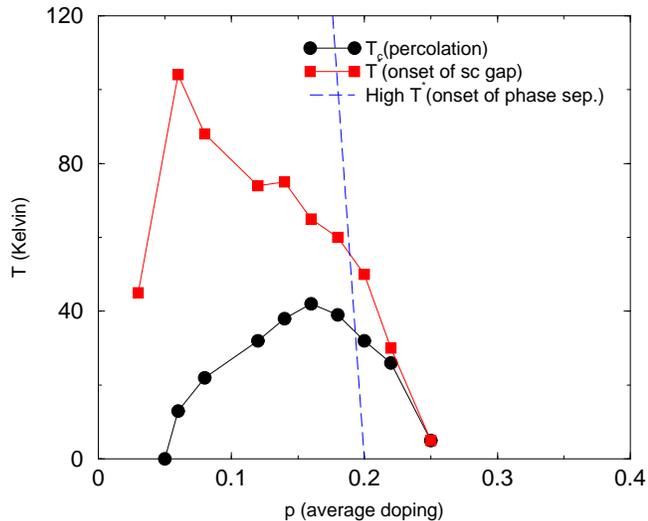}
\caption{(color online)
The onset of superconducting islands temperature $T^*(p_m)$ and
the percolation temperature $T_c(p_m)$ taken from Figs.(\ref{Deltan14.22T})
and (\ref{Deltan03.12T}).
It is also shown the phase separation (dashed) line or upper
pseudogap\cite{Tallon,Naqib}.}
\label{TempPhase}
%\end{center}
\end{figure}

In Fig.(\ref{TempPhase}) we used the calculated local $\Delta(i,T)$
on each  compound to
derive the LSCO  phase diagram, to wit, the onset of superconducting
temperature  $T^*(p_m)$
(squares) and the percolation temperature  $T_c(p_m)$(circles)
as function of $p_m$, as  derived from
Figs.(\ref{Deltan14.22T}) and (\ref{Deltan03.12T}).
The values of $T^*(p_m)$  are in good
agreement with the measurements attributed to the lower pseudogap
which is usually  related with the onset of a superconducting
property\cite{TS,Tallon,Markiewicz} as, for instance, the tunneling
results\cite{Renner,Mourachkine} and the Nernst
effect\cite{Ong}. The values of $T_c(p_m)$ is
also in good quantitative agreement with the experimental
superconducting phase boundary\cite{TS,Tallon,Ong}.

\begin{figure}[!ht]
\begin{center}
\includegraphics[height=7cm]{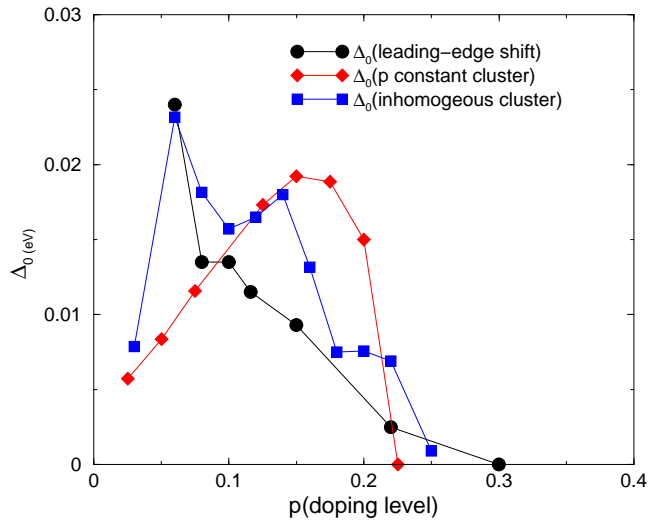}
\caption{ The zero temperature superconductor gap as function
of the doping level. The diamond points are for a cluster with
uniform density. The square are for an inhomogeneous cluster
according the CH results. The circles are the experimental leading
edge gap from Ref.(\cite{Ino}).}
\label{Delta0}
\end{center}
\end{figure}

We have also taken the larger superconducting gaps at low
temperatures for each compound from Figs.(\ref{Deltan03.12T}) and
(\ref{Deltan14.22T}) and plotted in Fig.(\ref{Delta0}). The
calculated maximum gap for each compound at low temperature
($\Delta_0(p_m)$) is in reasonable agreement with the ARPES zero
temperature leading edge shift or the magnitude of the
superconducting gap\cite{Ino}. In this figure, just to show the
effect of the disorder in our calculations, we have shown the values
of $\Delta_0(p_m)$ for homogeneous compounds. We see that the
disorder increases the average zero temperature gap dramatically at
low doping and in a weaker way, in the far overdoped regions. The
discrepancies around optimally doped samples between the
experimental values and our calculations are likely to be due to our
approximate stripe configurations on small clusters.

It is important to notice that this stripe scheme is able to capture
the very curious dual behavior of the electronic structure in LSCO
systems\cite{Zhou}. At the lightly doped regime, due to the high
values of the phase separation energy barrier  $E_g(T,p_m)$, the
charges move preferably along the high densities stripes, and the
Cooper pairs are formed along them, as demonstrated in
Figs.(\ref{Deltan03.12T}) and (\ref{Deltan14.22T}) by the calculated
$\Delta(i,T=0K)$. Consequently, with a k-space probe, the
superconducting gaps for lightly doped samples are measured mainly
in the $(\pi,0)$ and $(0,\pi)$ antinodal regions and the spectral
weight segments are entirely near these antinodal regions. This
behavior is expected for 1D stripes\cite{Zhou} and the measured
values of the zero temperature leading edge as function of $p_m$ can
be qualitatively reproduced by the $\Delta_0(p_m)$, as shown in
Fig.(\ref{Delta0}).

Since by assumption $E_g(T,p_m)$ is closely related with the high
pseudogap and, therefore decreases rapidly with $p_m$, it is easier
to the holes in compounds near optimally doped than underdoped ones
to tunnel over among the high and low doping stripes, yielding a 2D
character to these systems. Thus as $p_m$ increases, the samples
tend to change continuously from 1D to a 2D metallic behavior and
this is detected by the increase of the spectral weight near the
Fermi surface along the [1,1] nodal direction\cite{Zhou}. Another
evidence of the charge tunneling between different stripes is the
presence of the incoherent ZTPG measured on the surface of Bi2212
compounds\cite{McElroy1,Fang} which should scaled with the energy barrier
$E_g(p_m,T)$. The ZTPG are more frequently than the superconducting
gap in the underdoped region and disappear near
$p_m=0.19$\cite{McElroy1,Fang,McElroy2}. Evidence of these two types of
gaps in HTSC compounds came also from tunneling experiments carried
with different resistances\cite{Moura,Mourachkine}.

%\includegraphics[height=6cm]{Tgaussn022d04.eps}
%\caption{(color online)
%The temperature evolution of the local gap for a overdoped
%compound of $p=0.22$, in units of $eV$. It is shown just the first
%sites of a larger cluster with all the sites having this behavior.}
%\label{Tgauss}
%%\end{center}
%\end{figure}

\section{Conclusions}

We have worked out a complete scenario to HTSC and provided an
interpretation to the upper and lower pseudogap line and also to the
superconducting phase. Taking the upper pseudogap as the phase
separation temperature, we have calculated the local pairing
amplitudes $\Delta(i,T)$ which, as in BCS, is assumed to have their
phase locked and the superconducting phase is reached by percolation
at $T_c(p_m)$. In this way, we derived  the phase diagram of the
onset of $\Delta(i,T)$, that is $T^*(p_m)$ and $T_c(p_m)$ for a HTSC family. Although
the process of phase separation varies continuously with the
temperature and depends on the sample preparation, for simplicity,
we use in our calculations only the low temperature static
configuration. Despite this simplification, the method is quite
general, and to demonstrate it, we reproduced results in good
agreement with the LSCO series.

The many values of $\Delta(i,T)$ in a single compound agrees with
the several recent very fine STM maps on Bi2212 surfaces. The
calculations with the CH stripe configuration have also provided
novel interpretation to important aspects involved in the electronic
structure of this type of disorder: the inhomogeneous electronic
dual nature of underdoped LSCO cuprates measured by the straight
(1D) segments near the anti nodal regions and the spectral weight
near the nodal regions where the Fermi surface develops.  The
presence of the ZTPG peaks in several Bi2212 compounds, as measured
by high resolution STM data, may be due to the energy barrier
between the low and high density regions in a given sample and
therefore, in this view, is connected to the phase separation and
not with the superconductivity. Our results indicate that the normal
phase of cuprates is a disordered metal for $p_m\le 0.2$ composed by
the coexistence of insulator and metallic regions below the phase
separation temperature $T_{ps}$. These regions are composed of non
constant hole density regions which yields a variety of local
$T_c(p(i))$ at low temperatures. This inhomogeneity is the source of
scattering between these non-uniform regions, Andreev reflection,
etc, what is the cause of many nonconventional transport
properties\cite{TS,Lee}. The study of these normal phase properties will
be matter of future publication.

Thus, in short, with the
combination of the CH phase separation model
with the BdG superconducting theory, we have shown that the phase
diagram and some nonconventional properties of HTSC receive a
coherent interpretation based on a temperature dependent static
phase separation.

This work has been partially supported by CAPES, CNPq and CNPq-Faperj
Pronex E-26/171.168/2003.

\end{document}